\documentclass[NewProceedings,letterpaper]{ascelike-new}
\usepackage[utf8]{inputenc}
\usepackage[T1]{fontenc}
\usepackage{lmodern}
\usepackage{graphicx}
\usepackage[figurename=Fig.,labelfont=bf,labelsep=period]{caption}
\usepackage{subcaption}
\usepackage{amsmath}
\usepackage{amsfonts}
\usepackage{amssymb}
\usepackage{amsbsy}
\usepackage{newtxtext,newtxmath}
\usepackage[colorlinks=true,citecolor=black,linkcolor=black]{hyperref}

\newtheorem{lemma}{Lemma}

\def\reals{\mathbb{R}} 
\def\Gsn{\mathcal{N}}

\usepackage{color}
\newcommand{\rev}[1]{{\leavevmode\color{black}{#1}}}
\NameTag{Liao, October 1, 2017}
\newcommand{\mv}{\mathbf{x}}

\begin{document}

\title{Structural Damage Detection and Localization with Unknown Post-Damage Feature Distribution Using Sequential Change-Point Detection Method}

\author[1]{Yizheng Liao}
\author[2]{Anne S. Kiremidjian}
\author[3]{Ram Rajagopal}
\author[4]{Chin-Hsuing Loh}

\affil[1]{Ph.D. Candidate, Dept. of Civil and Environmental Engineering, Stanford University. 473 Via Ortega, Stanford, CA, 94305, USA. Email: yzliao@stanford.edu}
\affil[2]{Professor, Dept. of Civil and Environmental Engineering, Stanford University. 473 Via Ortega, Stanford, CA, 94305, USA. Email: ask@stanford.edu}
\affil[3]{Assistant Professor, Dept. of Civil and Environmental Engineering, Stanford University. 473 Via Ortega, Stanford, CA, 94305, USA. Email: ramr@stanford.edu}
\affil[4]{Professor, Dept. of Civil Engineering, National Taiwan University. No.1, Sec. 4, Roosevelt Road, Taipei, 106 Taiwan. Email: lohc0220@ntu.edu.tw}

\maketitle

\begin{abstract}
The high structural deficient rate poses serious risks to the operation of many bridges and buildings. To prevent critical damage and structural collapse, a quick structural health diagnosis tool is needed during normal operation or immediately after extreme events. In structural health monitoring (SHM), many existing works will have limited performance in the quick damage identification process because 1) the damage event needs to be identified with short delay and 2) the post-damage information is usually unavailable. To address these drawbacks, we propose a new damage detection and localization approach based on stochastic time series analysis. Specifically, the damage sensitive features are extracted from vibration signals and follow different distributions before and after a damage event. Hence, we use the optimal change point detection theory to find damage occurrence time. As the existing change point detectors require the post-damage feature distribution, which is unavailable in SHM, we propose a maximum likelihood method to learn the distribution parameters from the time-series data. The proposed damage detection using estimated parameters also achieves the optimal performance. Also, we utilize the detection results to find damage location without any further computation. Validation results show highly accurate damage identification in American Society of Civil Engineers benchmark structure and two shake table experiments.
\end{abstract}

\section{Introduction}
As surveyed by American Society of Civil Engineers (ASCE), many structures in the U.S. are structurally deficient or in poor condition. For example, about $10\%$ bridges in the U.S., which carry $188$ million daily trips, are structurally deficient \cite{asce2017report}. $24\%$ of American public school buildings are in fair or poor condition. To avoid the accident like the collapse of I-35 Mississippi River bridge in 2007, it is necessary to continuously monitor the performance level and safety of structures and provide quick assessment of the severity of damage during daily operation and immediately after extreme events like earthquakes and hurricanes.

During the past several decades, the statistical pattern recognition (SPR) has received significant attentions in the field of structural health monitoring (SHM), especially in the context of vibration analysis of structures \cite{farrar1999statistical,farrar2000pattern}. In SPR paradigm, the changes in the physical structures, such as a loss of stiffness, lead to the changes in the sensor measurements and the damage sensitive features (DSFs), which are extracted from the acquired data. Therefore, damage can be detected through changes or outliers in the DSFs rather than changes in the structural properties. As a result, the knowledge of the structural properties is no required in SPR. Many existing works have used SPR methods to detect and localize damage, including \cite{nair2006time,yeung2005damage,bornn2015modeling,qiao2012signal,lei2003statistical,huang2017hierarchical,hui2017structural}.

However, many of these existing approaches will have limit performance in rapid structural health assessment. For example, \cite{hui2017structural} and \cite{lanata2006damage} need to collect structural responses over a long period of time to achieve accurate damage detection. Unfortunately, this requirement introduces a significant detection delay in the damage diagnosis process. Also, some methods, such as \cite{tian2014damage} and \cite{yeung2005damage}, require the post-damage information, such as post-damage feature distribution. This information is unavailable in many applications due to the complexity of structures and loading conditions. Even it is available, this information may be incorrect or does not cover all possible damage patterns. Furthermore, many works only focus on damage detection but not damage localization, which is an important component of SHM and can significantly reduce the time of structural health diagnosis.

In order to tackle the challenges of quick structural health diagnosis, building on our previous work \cite{noh2013sequential}, we propose a sequential damage detection and localization algorithm using the vibration responses from multiple sensors. In particular, the acceleration measurements obtained from sensors are modeled as an autoregressive (AR) time series. The damage sensitive features (DSFs) are defined as a function of the AR coefficients. It has been found that the DSF distribution parameters for the damaged and undamaged signals are different \cite{nair2006time}. A well-known method to sequentially detect the probability distribution change is the change point detection method \cite{tartakovsky2005general}. In change point detection, the detector observes a sequence of DSFs and reports a damage event when it detects a change of DSF probability distribution due to some events at an unknown time. The objective is identifying a damage event as quickly as possible subject to a fixed probability of false alarm. As discussed previously, the post-damage distribution is hard to obtain. To overcome this drawback, we propose a maximum likelihood method to estimate the unknown distribution parameters from data directly. Unlike our previous work in \cite{noh2013sequential}, the proposed estimator can apply to multivariate distributions and does not require the DSF distributions to have the same covariance before and after damage events. The damage detection is performed at each sensor. Hence, we propose two damage localization indices that utilizing the sensor topology information and damage detection results. We validate the proposed damage identification method by numerical simulation of the ASCE benchmark structure with various damage patterns and the data sets of two shake table experiments.

The rest paper is organized as follows: the Sequential Damage Detection Algorithm section introduces the sequential damage diagnosis process, defines the DSF, and proposes the sequential damage detection algorithm. Also, a maximum likelihood method is presented to estimate post-damage DFS distribution parameters. The Structural Damage Localization Indices section presents two damage localization indices utilizing damage detection results and sensors' location information. The Results of Simulation and Experiment section presents the validation results of the proposed damage identification algorithm on data sets collected from simulation and shake table experiments. The last section concludes this paper.

\section{Sequential Damage Detection Algorithm}
\label{sec:alg_detect}

\begin{figure}
\centering
\includegraphics[width=0.8\linewidth]{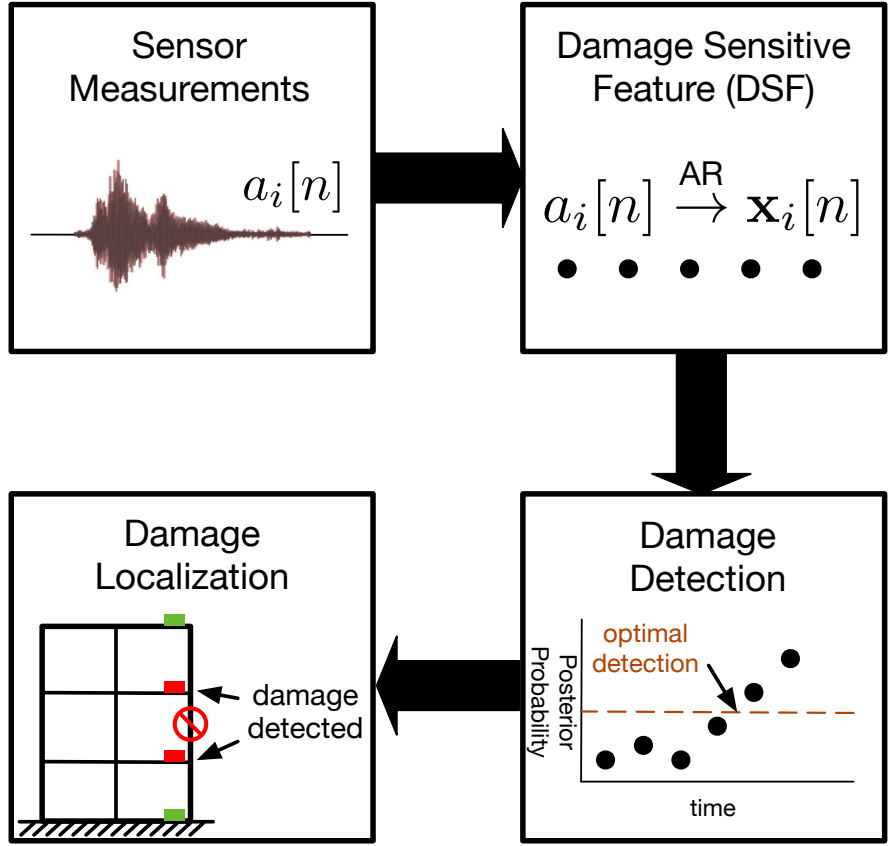}
\caption{Summary of the damage diagnosis process.}
\label{fig:process}
\end{figure}

The proposed sequential damage detection and localization algorithm (Fig.~\ref{fig:process}) has four steps: (i) structural response acquisition, (ii) DSF extraction, (iii) damage detection, and (iv) damage localization. In the first step, the structural responses are sequentially obtained from the monitoring devices, such as accelerometers and gyroscopes. Then, we extract the DSFs from these collected signals. The DSF is required to have statistical changes before and after damage occurrence. In this study, we use the coefficients of acceleration AR model as an example of DSF. \cite{nair2006time} has shown that the AR model coefficients extracted from the acceleration signals are related to structural parameters and sensitive to structural damage. We want to highlight that the proposed algorithm also works for other types of DSFs, such as wavelet coefficients of acceleration signal \cite{nair2009derivation}, AR model coefficients of angular velocity \cite{liao2016angular}. The DSF extraction consists of two steps: (i) normalization and (ii) AR model fitting. The discrete time acceleration signal from sensor $i$, $a_i[n]$, is divided into chunks with a size $M$. Let $a_i^k[n]$ denote the $k$-th chunk of the signal $a_i[n]$. The signal is normalized as $\widetilde{a}_i^k[n] = (a_i^k[n]-\mu^k_i)/\sigma^k_i$, where $\mu^k_i$ and $\sigma^k_i$ denote the mean and standard deviation of the $k$-th chunk data. The normalized data are fitted with a single-variate AR model of order $p$, i.e.,
\begin{equation}
\label{eq:AR}
\widetilde{a}_i^k[n] = \sum_{j=1}^p \gamma_j^k \widetilde{a}_i^k[n-j] + \epsilon_i^k[n],	
\end{equation}
where $\gamma_j^k$ denotes the $j$-th AR coefficient and $\epsilon_i^k[n]$ denotes the residual. The process of model order selection has been discussed with details in \cite{noh2013sequential}. For connivence, we define $\mathbf{x}_i[n] \in \reals^{m}$ as the DSF of sensor $i$ extracted from the $n$-th chunk. $\mathbf{x}_i[n]$ is defined as a vector because it may contain multiple AR coefficients, e.g., $\mathbf{x}_i[n] = [\gamma_1^n, \gamma_2^n]^T$. Since we collect structural responses continually, we will have a time-series of DSFs, e.g., $\mathbf{x}_i^N = \{\mathbf{x}_i[1],\mathbf{x}_i[2], \cdots, \mathbf{x}_i[N]\}$.

In the third step of the damage diagnosis process, a statistical test is conducted on the extracted DSFs sequentially. Once a damage event is detected, we use the statistics of DSFs and sensors' location to find damage location, as shown in the fourth step of Fig.~\ref{fig:process}. Since the DSF is a time-series signal, one way to represent the time-series data is using random variables. Therefore, we model the DSF at sensor $i$ as a random vector $\mathbf{X}_i$ and $\mathbf{x}_i[n]$ is the realization of $\mathbf{X}_i$ at time step $n$. As discussed in \cite{nair2007time}, the AR coefficient-based DSF $\mathbf{X}_i$ follows a multivariate Gaussian distribution and its mean and covariance change after a damage event. As a result, the probability distributions of $\mathbf{X}_i$ is different before and after damage occurrence. Let $\lambda$ denote the time of damage occurrence. We assume that $\mathbf{X}_i$ follow $\Gsn(\boldsymbol{\mu}_0,\Sigma_0)$ in the normal operation (i.e., $N < \lambda$), and the other Gaussian distribution $\Gsn(\boldsymbol{\mu}_1,\Sigma_1)$ after any damage event (i.e., $N \geq \lambda$). Finding the damage occurrence time $\lambda$ sequentially is equivalent to perform the following hypothesis test at each time $N$:
\begin{eqnarray*}
	\text{Pre-damage} \quad & \mathcal{H}_0: & \lambda > N, \\
	\text{Post-damage} \quad & \mathcal{H}_1: & \lambda \leq N.
\end{eqnarray*}
A well-known approach for solving this sequential hypothesis testing problem is the change point detection method \cite{tartakovsky2005general}. Specifically, if we assume the damage occurrence time $\lambda$ follow a geometric distribution with a parameter $\rho$, i.e., $\lambda \sim Geo(\rho)$, we can use a Bayesian approach to find $\lambda$. The joint distribution of $\lambda$ and $\mathbf{X}_i$ can be written as $P(\lambda,\mathbf{X}_i) = \pi(\lambda) P(\mathbf{X}_i|\lambda)$, where $\pi(\lambda)$ is the probability mass function of $\lambda$. When $\lambda = k$, all the data obtained before time $k$ follow the pre-damage distribution $g(\mathbf{X})$ and all the data obtained after time $k$ follow the post-damage distribution $f(\mathbf{X})$, which can be further expressed as:
\begin{equation}
\label{eq:likelihood}
P(\mathbf{X}_i = \mathbf{x}_i^N|\lambda=k) = \prod_{n=1}^{k-1}g(\mathbf{x}_i[n])\prod_{n=k}^{N}f(\mathbf{x}_i[n]),
\end{equation}
for $k = 1,2,\cdots,N+1$. At $\lambda = N+1$, the damage has not occurred and all data follow $g(\mathbf{X})$.

In order to find the damage occurrence time $\lambda$, we compute the post-damage posterior probability $P(\mathcal{H}_1|\mathbf{X}_i) = P(\lambda \leq N | \mathbf{X}_i = \mathbf{x}_i^N)$. At each time $N$, 
\begin{eqnarray}
	P(\lambda \leq N | \mathbf{x}_i^N) &=& \sum_{k=1}^N \frac{P(\lambda = k, \mathbf{x}_i^N)}{P(\mathbf{x}_i^N)} = \frac{1}{P(\mathbf{x}_i^N)}\sum_{k=1}^N \pi(\lambda=k)P(\mathbf{x}_i^N|\lambda=k) \nonumber \\
	&=& C\sum_{k=1}^N\pi(k)\prod_{n=1}^{k-1}g(\mv_i[n])\prod_{n=k}^Nf(\mv_i[n]), \label{eq:post_prob}
\end{eqnarray}
where $C$ is a normalization factor such that $\sum_{k=1}^{N+1}P(\lambda=k | \mathbf{x}_i^N) = 1$. In the normal operation, $f(\mv_i[n])$ is small and $P(\lambda \leq N | \mathbf{x}_i^N)$ is small. Once a damage event occurs at time $\lambda=k \leq N$, all data collected at $n \geq \lambda$ follow $f(\mathbf{X})$ and hence, $P(\lambda \leq N | \mathbf{x}_i^N)$ becomes large. To detect the damage time $\lambda$, we can set a threshold and declare the damage event when the post-damage posterior probability surpasses this threshold. 

\subsection{Optimal Damage Detection}
In the change point detection problem, we are interested in two performance metrics: \textit{probability of false alarm} and \textit{average detection delay}. The former metric is the probability that a detector falsely declares a damage event in the normal operation. Letting $\tau$ denote the damage detection time, the probability of false alarm is defined as $P(\tau < \lambda)$. The latter metric describes the average latency that a detector finds the damage after it has occurred. The average detection delay is defined as $E(\tau - \lambda | \tau \geq \lambda)$. In structural damage detection problem, we want to detect the damage happens at time $\lambda$ \rev{as quickly as possible} under a constraint of the maximum probability of false alarm $\alpha$, i.e.,
\[
	\underset{\tau}{\text{minimize}} \quad E(\tau - \lambda | \tau \geq \lambda) \quad\quad \text{subject to} \quad P(\tau < \lambda) \leq \alpha.
\]
By the Shiryaev-Roberts-Pollaks procedure \cite{pollak2009optimality}, given all observed DSFs $\mv_i^N$, the optimal solution for the optimization above is \cite{tartakovsky2008asymptotically}
\begin{equation}
	\label{eq:detect_rule}
	\tau = \inf\{N \geq 1:P(\lambda \leq N | \mathbf{x}_i^N) \geq 1-\alpha\}.
\end{equation}
In (\ref{eq:detect_rule}), if we fix $\alpha$ in advance, the threshold will be the same for all detectors. With this detection rule, the average detection delay is presented in Lemma~\ref{thm:opt_delay}.

\begin{lemma}
	\label{thm:opt_delay}
	For a given probability of false alarm $\alpha$, the detection rule in (\ref{eq:detect_rule}) achieves the asymptotically optimal detection delay
	\begin{equation}
	D(\tau) = E(\tau - \lambda|\tau \geq \lambda) = \frac{|\log\alpha|}{-\log(1-\rho)+D_\text{KL}(f(\mathbf{X})\|g(\mathbf{X}))},
	\end{equation}
	as $\alpha \rightarrow 0$, where $\rho$ is the prior distribution parameter and $D_\text{KL}(f(\mathbf{X})\|g(\mathbf{X}))$ is the Kullback-Leibler distance \cite{tartakovsky2008asymptotically}.
\end{lemma}
In Lemma~\ref{thm:opt_delay}, if $\rho$ and $D_{KL}(f(\mathbf{X})\|g(\mathbf{X}))$ are constants, the detection delay $D(\tau)$ is longer if $\alpha$ is small and vice versa. Therefore, in the field applications, we need to decide the trade-off between the detection delay and the probability of false alarm.

In summary, when a new group of DSF $\mv_i[n]$ is available, we compute the post-damage posterior probability according to (\ref{eq:post_prob}) and then apply the optimal detection rule in (\ref{eq:detect_rule}) to assess the structural health condition. Since we process the data sequentially, our method can provide real-time structural health assessment information for civil engineers and structure managers. Also, the proposed approach allows us to perform the structural health diagnosis process on the sensing device and only transmit detection results, e.g., damage detection time. This benefit minimizes the wireless communication in wireless sensor networks (WSNs), which is the most power consuming activity in WSN \cite{liao2014snowfort}, and extend the lifetime of wireless sensors.

\subsection{Damage Detection with Unknown Post-Damage Feature Distribution}

The computation of $P(\lambda \leq N | \mathbf{x}_i^N)$ in (\ref{eq:post_prob}) requires the knowledge of both $g(\mathbf{X})$ and $f(\mathbf{X})$. For the pre-damage DSF distribution $g(\mathbf{X})$, we can learn the parameters $\{\boldsymbol{\mu}_0$, $\Sigma_0\}$ from historical data collected during the normal operation. However, the parameters of $f(\mathbf{X})$, $\boldsymbol{\mu}_1$ and $\Sigma_1$, are hard to obtain because the damage pattern is usually unpredictable. With the information on structure and materials, experts may create a database to contain every possible damage pattern and use the observed DSFs to search the most likely damage pattern. However, this approach is infeasible for a large-scale structure. For example, the structure in Fig.~\ref{fig:asce} has 32 braces. If a damage pattern only involves single brace broken, there are 32 damage patterns. If two or more braces are damaged at once, the possible damage patterns will increase exponentially. 

In this section, instead of searching the most likely damage pattern, we propose an algorithm to learn $f(\mathbf{X})$ from data using maximum likelihood method. The computational complexity of our approach is insensitive to the complexity of structure. To apply the maximum likelihood method, we need to set $\partial P(\lambda \leq N | \mathbf{x}_i^N)/\partial \boldsymbol{\mu}_1 = 0$ and $\partial P(\lambda \leq N | \mathbf{x}_i^N)/\partial \Sigma_1 = 0$. Unfortunately, $P(\lambda \leq N | \mathbf{x}_i^N)$ is not a convex function and we may have multiple estimates. To address this drawback, we provide an approximation of $P(\lambda \leq N | \mathbf{x}_i^N)$. Specifically, the log-probability $\log P(\lambda \leq N | \mathbf{x}_i^N)$ can be expressed as
\begin{equation}
\label{eq:log_post_prob}	
\log P(\lambda \leq N | \mathbf{x}_i^N) = \log C + \log\left\{\sum_{k=1}^N\pi(k)\prod_{n=1}^{k-1}g(\mv_i[n])\prod_{n=k}^Nf(\mv_i[n],\boldsymbol{\theta})\right\},
\end{equation}
where $\boldsymbol{\theta} = \{\boldsymbol{\mu}_1,\Sigma_1\}$ represents the unknown parameters of $f(\mathbf{X})$. In (\ref{eq:log_post_prob}), the term within the braces can be considered as an expectation of $\prod_{n=1}^{k-1}g(\mv_i[n])\prod_{n=k}^Nf(\mv_i[n],\boldsymbol{\theta})$ over the prior distribution $\pi$, i.e., $E_\pi\left(\prod_{n=1}^{k-1}g(\mv_i[n])\prod_{n=k}^Nf(\mv_i[n],\boldsymbol{\theta})\right)$. Also, the logarithmic function is convex. Thus, we can apply the Jensen's inequality \cite{cover2012elements} to approximate $\log P(\lambda \leq N | \mathbf{x}_i^N)$, i.e.,
\begin{equation}
	\label{eq:approx_log_post_prob}
\log P(\lambda \leq N | \mathbf{x}_i^N) \geq  \log C + \sum_{k=1}^N\pi(k) \left(\sum_{n=1}^{k-1}\log(g(\mv_i[n])) + \sum_{n=k}^{N}\log(f(\mv_i[n],\boldsymbol{\theta})) \right) \triangleq  \widetilde{P}(\lambda \leq N | \mathbf{x}_i^N). 
\end{equation}
Since $\widetilde{P}(\lambda \leq N | \mathbf{x}_i^N)$ is convex, by setting $\partial \widetilde{P}(\lambda \leq N | \mathbf{x}_i^N)/\partial \boldsymbol{\mu}_1 = 0$ and $\partial \widetilde{P}(\lambda \leq N | \mathbf{x}_i^N)/\partial \Sigma_1 = 0$, the parameter $\boldsymbol{\theta}$ can be estimated as
\begin{equation}
	\label{eq:parm_est}
	\widehat{\boldsymbol{\mu}}_1 = \frac{\sum_{k=1}^N \pi(k)\sum_{n=k}^N\mv_i[n]}{\sum_{k=1}^N \pi(k) (N-k+1)}, \quad
	\widehat{\Sigma}_1 = \frac{\sum_{k=1}^N \pi(k)\sum_{n=k}^N(\mv_i[n]-\widehat{\boldsymbol{\mu}}_1)(\mv_i[n]-\widehat{\boldsymbol{\mu}}_1)^T}{\sum_{k=1}^N \pi(k) (N-k+1)}.
\end{equation}
The details of (\ref{eq:parm_est}) are provided in Appendix~\ref{sec:par_est}. Now, we can use $\widehat{\boldsymbol{\mu}}_1$ and $\widehat{\Sigma}_1$ to compute $P(\lambda \leq N | \mathbf{x}_i^N)$ and then apply the optimal detection rule in (\ref{eq:detect_rule}).

\section{Structural Damage Localization Indices}
\label{sec:alg_local}
Identifying the damaged component or finding the damage location is important in SHM. An efficient and accurate damage localization approach can significantly reduce the time of structural health diagnosis. In this section, we will propose two real-time damage localization methods based on the DSFs collected at each sensor and sensors' location.

\begin{figure}
\centering
\includegraphics[width=\linewidth]{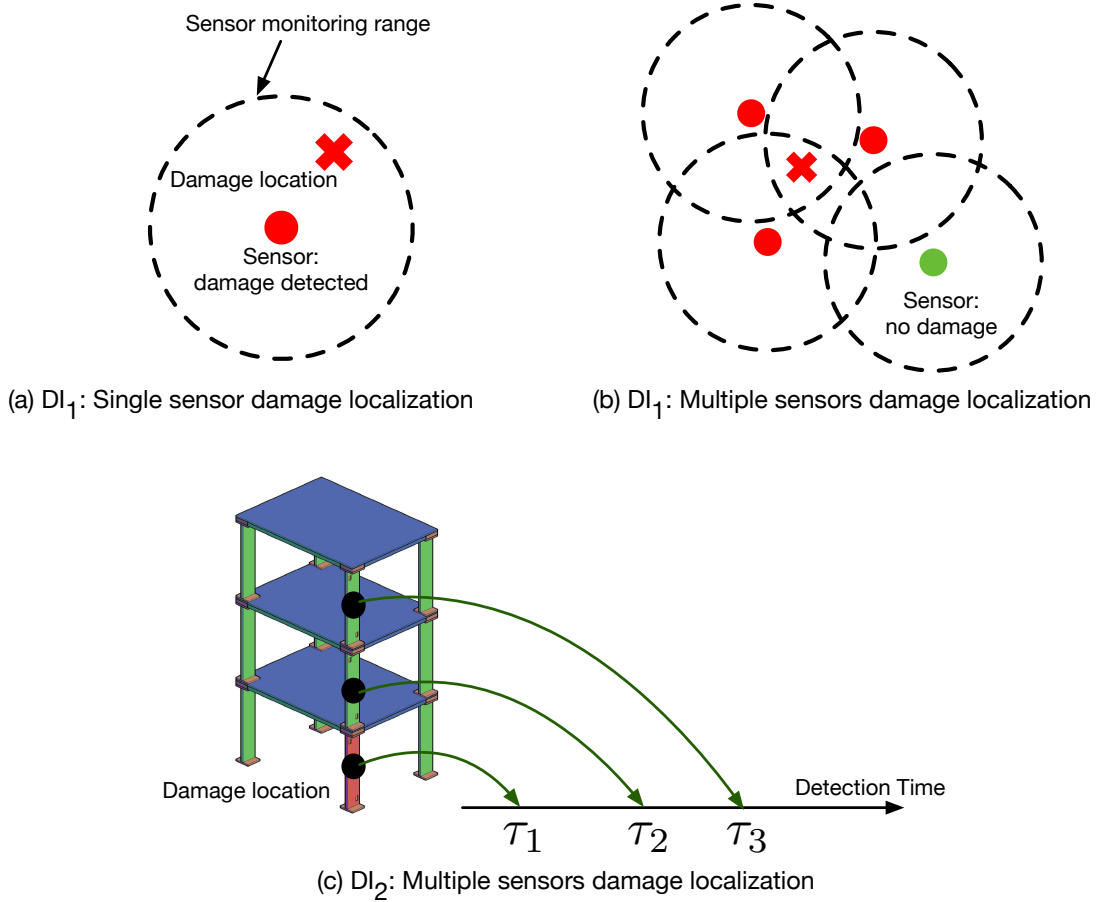}
\caption{Examples of damage localization indices.}
\label{fig:DI_example}	
\end{figure}

When a damage event occurs, the DSFs collected from sensors that near the damage location have significant changes. In Information Theory, a widely used metric to describe the difference between two distributions is the Kullback-Leibler (KL) distance \cite{cover2012elements}, which is defined as $D_{KL}(f(\mathbf{X})\|g(\mathbf{X})) = \int_{-\infty}^\infty f(\mathbf{x})\log(f(\mathbf{x})/g(\mathbf{x}))d\mathbf{x} \geq 0$. When two distributions are identical, the KL distance is zero. For the sensor near the damage location, the KL distance between pre-damage DSF distribution $g(\mathbf{X})$ and post-damage DSF distribution $f(\mathbf{X})$ is large. Therefore, we can use the KL distance computed at each sensor location to narrow down the possible damage areas, as demonstrated in Fig.~\ref{fig:DI_example}. For the multivariate Gaussian distribution, the KL distance can be computed as
\begin{equation}
\label{eq:KL}
	D_{KL}(f(\mathbf{X})\|g(\mathbf{X})) = \frac{1}{2}\left\{(\text{tr}(\Sigma_0^{-1}\Sigma_1) + (\boldsymbol{\mu}_0 - \boldsymbol{\mu}_1)^T\Sigma_0^{-1}(\boldsymbol{\mu}_0 - \boldsymbol{\mu}_1) - m + \ln\left(\frac{\text{det}(\Sigma_0)}{\text{det}(\Sigma_1)}\right)\right\},
\end{equation}
where $m$ the dimension of $\mathbf{X}_i$, the operator $\text{tr}()$ is the matrix trace operator, and $\text{det}()$ is the matrix determinant operator. The computation of (\ref{eq:KL}) requires the parameters of $g(\mathbf{X})$ and $f(\mathbf{X})$. Although the parameters of $f(\mathbf{X})$ are usually unknown, fortunately, we can use the estimated parameters in (\ref{eq:parm_est}) for computation. Therefore, we can use the KL distance to localize damage, e.g., $\text{DI}_1 = D_{KL}(f(\mathbf{X})\|g(\mathbf{X}))$, with or without knowing $f(\mathbf{X})$.

As discussed earlier, the lifetime of battery is one of the key bottlenecks that prevent the long-time WSN deployment. Although we have reduced power consumption by utilizing onboard computation capability, we can further extend the lifetime by minimizing the onboard computation. In (\ref{eq:KL}), the matrix multiplication and inversion have high computational complexity. To avoid these computations, we propose the second damage localization index based on the damage detection time. Specifically, with a given $\alpha$ and $\rho$, the optimal detection delay ($D(\tau)$) in Lemma.~\ref{thm:opt_delay} is a function of $D_{KL}(f(\mathbf{X})\|g(\mathbf{X}))$. If a sensor is installed near the damage location, $D_{KL}(f(\mathbf{X})\|g(\mathbf{X}))$ is large and the sensor reports damage faster than other sensors that are far away from the damage location, as shown in Fig.~\ref{fig:DI_example}. Therefore, we can use the damage detection time $\tau$ as the localization index, i.e., $\text{DI}_2 = \tau$. Compared with $\text{DI}_1$, $\text{DI}_2$ directly utilizes the damage detection results and requires no further computation. In next section, we will validate both localization indices using simulated and experimental data.




\section{Results of Simulation and Experiment}
\label{sec:sim}
In this section, we first validate the proposed damage detection and localization algorithm using the numerical simulation data from the ASCE benchmark structure \cite{johnson2004phase} with various damage patterns. Then we apply our approach to the data collected from two indoor shake table experiments. In all simulations and experiments, we assign an uninformative parameter for the prior distribution, e.g., $\rho = 10^{-5}$, and set the probability of false alarm as $\alpha = 10^{-5}$.

\subsection{ASCE Benchmark Structure Simulation}


\begin{figure}
\centering
\includegraphics[width=0.6\linewidth]{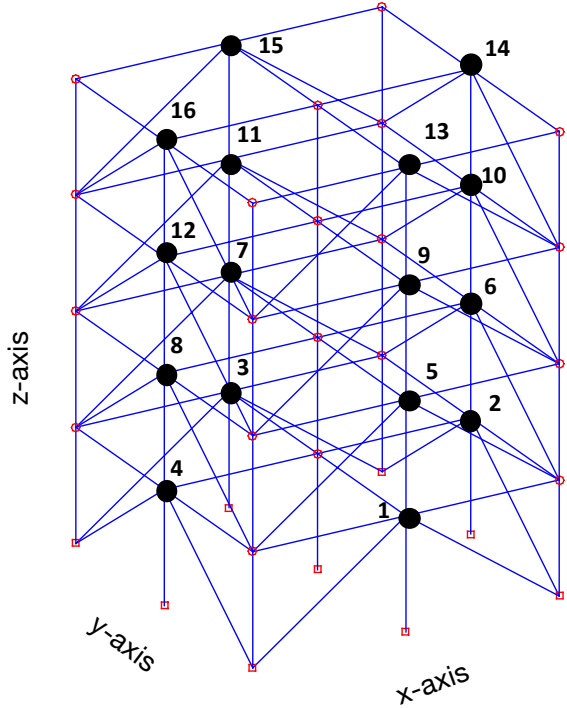}
\caption{ASCE benchmark structure with sensors (black dot).}
\label{fig:asce}
\end{figure}

\begin{figure}
\centering
\includegraphics[width=0.6\linewidth]{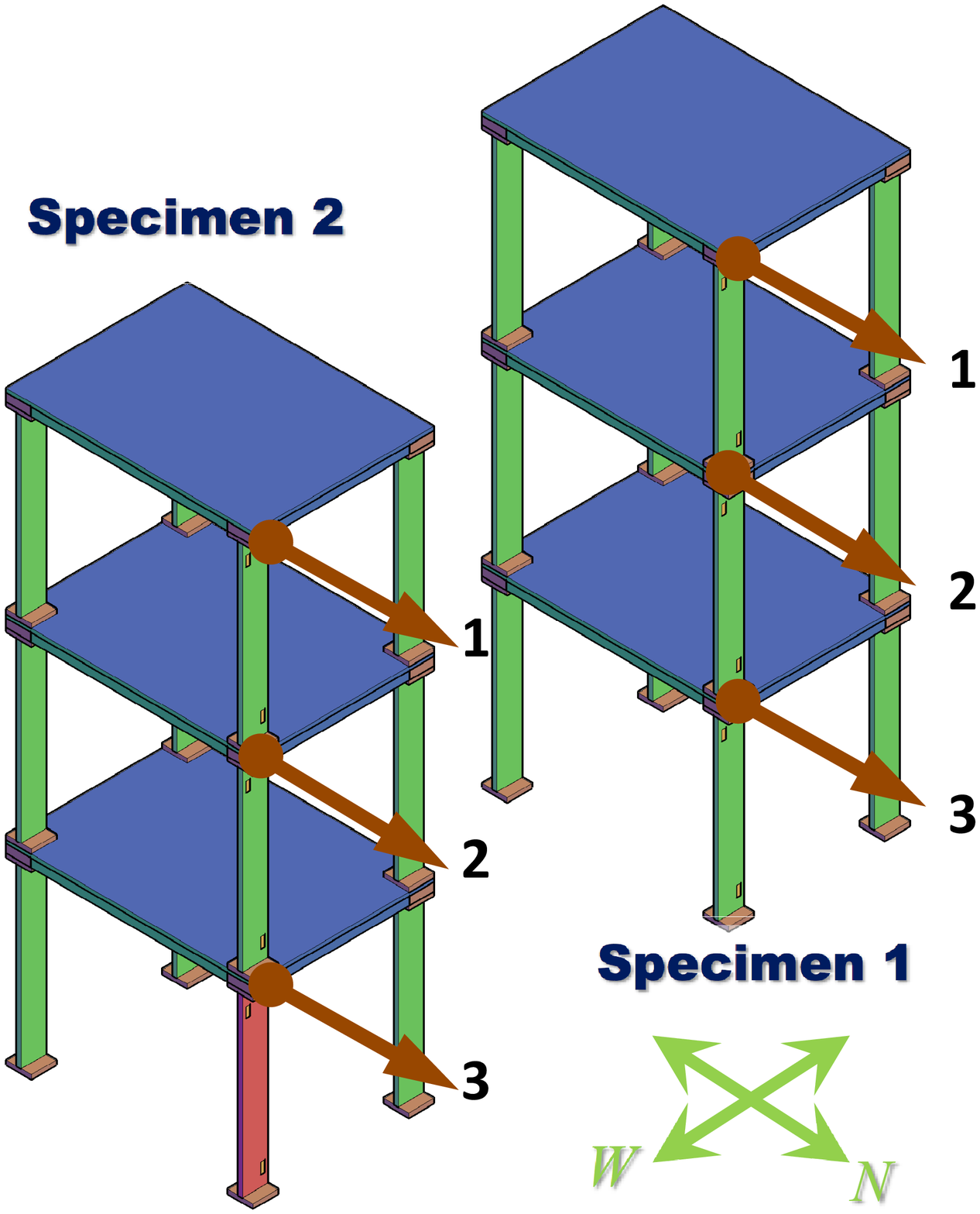}
\caption{Experiment 1: the diagram of structures and sensors.}
\label{fig:taiwan1}
\end{figure}

\begin{figure}
\centering
\includegraphics[width=\linewidth]{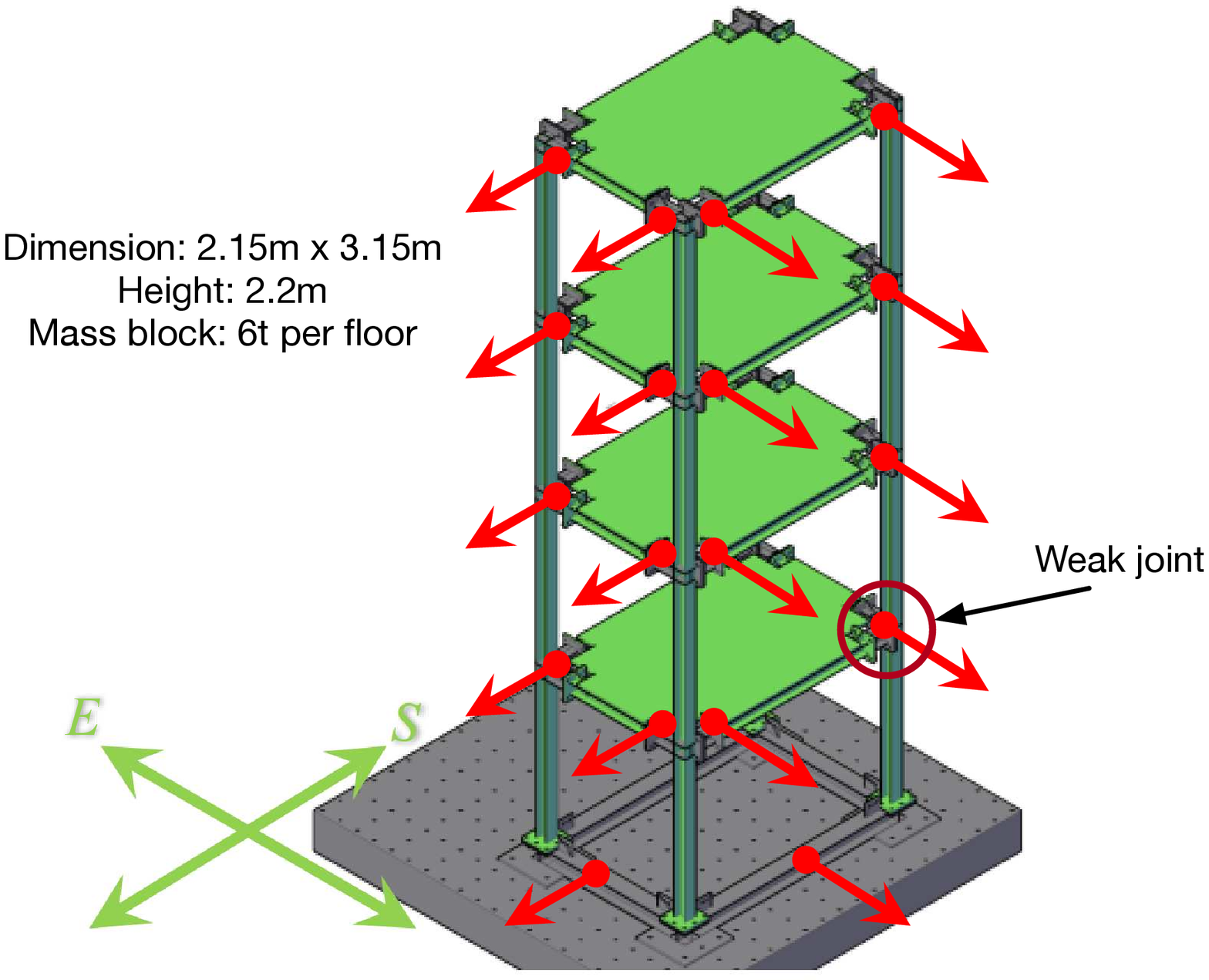}
\caption{Experiment 2: the diagram of structures and sensors.}
\label{fig:taiwan2}
\end{figure}

The ASCE benchmark structure is a four-story, two-bay by two-bay steel braced frame, as shown in Fig.~\ref{fig:asce}. The details about this structure and the simulator are provided in \cite{johnson2004phase}. The white noise excitations are applied on each floor along the y-axis. Damage pattern (DP) is simulated by removing braces in various combinations, including
\begin{itemize}
	\item DP 0: No damage
	\item DP 1: Removal of all braces below 1st floor;
	\item DP 2: Removal of all braces between 1st and 2nd floors;
	\item DP 3: Removal of all braces between 2nd and 3rd floors;
	\item DP 4: Removal of all braces between 3rd and 4th floors.
\end{itemize}

\begin{figure}
\centering
\includegraphics[width=\linewidth]{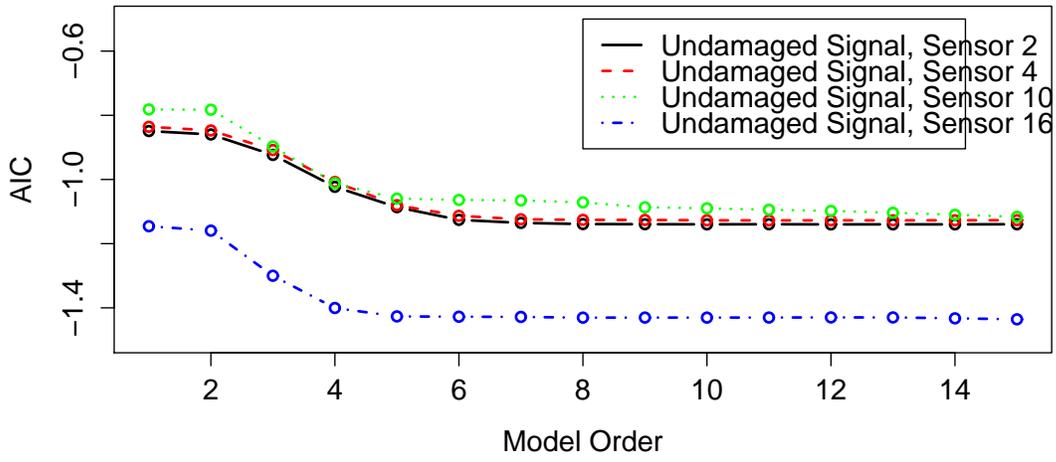}
\caption{\rev{AIC values against AR model order for the undamaged signals of ASCE benchmark structures.}}
\label{fig:AR_AIC}	
\end{figure}

In this simulation, four accelerometers are installed on each floor to measure the vibration along the y-axis. The structural responses are segmented into chunks and normalized, following the process described in the algorithm section. \rev{The normalized signals are fitted with an AR model. To choose the optimal order of AR model, we use the Akaike information criteria (AIC) \cite{hastie2009elements}. Fig.~\ref{fig:AR_AIC} shows the AIC values against AR model order for the undamaged signals. We observe that when $p=7$, the AIC values are steady. Therefore, we choose $p=7$ as the optimal values of AR model. A highlight is that we only choose the AR model order using the undamaged signal. As we discussed previously, the damage case is usually unknown in advance. Therefore, it is difficult to choose the AR model order of the damaged signals. As we shown below, using the same order of AR model for pre-damage and post-damage data does not affect the detection performance.}

All AR coefficients are used as the DSF, i.e., $\mathbf{X}_i = [\gamma_1,\gamma_2,\cdots,\gamma_7]^T$. The DSFs collected from DP 0 are used to estimate the parameters of $g(\mathbf{X})$. In this study, we only simulate one damage pattern every time. The damage occurrence time is $\lambda = 41$.
\begin{table}[h!]
\centering
\caption{ASCE benchmark structure: results of detection delay ($\tau-\lambda$).}
\label{tab:asce_delay}
\begin{tabular}{|c|c|c|c|c|c|c|c|c|}
\hline
\multicolumn{1}{|l|}{} & \multicolumn{4}{c|}{\begin{tabular}[c]{@{}c@{}}Detection Delay \\ (known parameters)\end{tabular}} & \multicolumn{4}{c|}{\begin{tabular}[c]{@{}c@{}}Detection Delay\\  (unknown parameters)\end{tabular}} \\ \hline
        Damage             & Sensor                & Sensor               & Sensor               & Sensor               & Sensor                & Sensor                & Sensor                & Sensor               \\
        Pattern & 2 & 6 & 10 & 14 & 2 & 6 & 10 & 14  \\ \hline
DP 1                   & 5                       & 10                     & 30                     & 38                     & 6                       & 13                      & 21                      & 36                     \\ \hline
DP 2                   & 0                       & 1                      & 26                     & 44                     & 3                       & 1                       & 26                      & 43                     \\ \hline
DP 3                   & 16                      & 0                      & 0                      & 14                     & 15                      & 4                       & 2                       & 14                     \\ \hline
DP 4                   & 24                      & 17                     & 0                      & 0                      & 12                      & 13                      & -1                      & 2                      \\ \hline
\end{tabular}
\end{table}

Table.~\ref{tab:asce_delay} shows the detection delays ($\tau-\lambda$). Since the structure is symmetric before and after damage, we only select one sensor from each floor. When the parameters of $f(\mathbf{X})$ are known, the sensors close to the damage location detect damage with a short delay. For example, for DP 3, Sensor 6 and Sensor 10, which are below and above the removed braces on the 3rd floor, detect damage with no delay. These results show that the detection time ($\text{DI}_2$) can serve as the damage localization index. When the parameters of $f(\mathbf{X})$ are unknown, the proposed method needs more time to detect damage. For example, Sensor 10 needs two more time steps to detect DP 3. In this simulation, the DSF is extracted every 2 seconds. Thus, the additional delay is only 4 seconds. Also,  the sensors that are close to the damage location still have the shortest detection delay. These results further validate $\text{DI}_2$. Sensor 10 has a negative delay in DP 4 due to a false detection. We can avoid this error by reducing the false alarm rate $\alpha$. In summary, when the parameters of $f(\mathbf{X})$ are unknown, our proposed method still has short detection delay and finds the damage locations successfully.

\begin{table}[h!]
	\centering
	\caption{ASCE benchmark structure: results of $\text{DI}_1$ with unknown parameters of $f(\mathbf{X})$.}
	\label{tab:asce_local}
	\begin{tabular}{|c||c|c|c|c|}
	\hline
	Sensor & DP 1 & DP 2 & DP 3 & DP 4 \\
	\hline
1 & 3.6014 & 44.7897 & 1.3007  & 0.5092 \\
2 & 4.2847 & 53.1589 & 1.4650  & 0.7484 \\
3 & 4.1470 & 47.1736 & 1.3183  & 0.4936 \\
4 & 3.2395 & 44.4338 & 1.0345  & 0.5107 \\
5 & 1.2482 & 23.0924 & 30.3441 & 0.4584 \\
6 & 1.5786 & 24.8039 & 33.0383 & 0.7293 \\
7 & 1.3294 & 23.4197 & 30.8325 & 0.5614 \\
8 & 1.3598 & 23.5146 & 31.2999 & 0.6227 \\
9  & 0.3759 & 0.7606 & 42.9107  & 72.7552 \\
10 & 0.4055 & 0.8413 & 49.1194 &  83.0035 \\
11 & 0.5778 & 0.8818 & 45.1768 &  74.4010 \\
12 & 0.4215 & 0.7138 & 48.3173 &  80.0961 \\
13 & 0.5347 & 0.6666 & 1.4975  &  99.2600 \\
14 & 0.6000 & 0.6162 & 1.6517  &  97.6008 \\
15 & 0.5910 & 0.5593 & 1.6600  &  94.8470 \\
16 & 0.5412 & 0.5313 & 1.6525  &  95.7695 \\
	\hline	
	\end{tabular}
\end{table}

Table.~\ref{tab:asce_local} shows the KL distance ($\text{DI}_1=D_{KL}(f(\mathbf{X})\|g(\mathbf{X}))$) when the parameters of $f(\mathbf{X})$ are unknown. The mean and covariance matrix are estimated using (\ref{eq:parm_est}). For DP 1, sensors on the first floor (Sensor 1-4) have relatively large $D_{KL}(f(\mathbf{X})\|g(\mathbf{X}))$. For DP 2, the KL distances of sensors on the first and second floors (Sensor 1-8) are significantly large and indicate that damage occurs between the first and second floors. Similar performances are observed in other DPs. These results show that $\text{DI}_1$ is sensitive to damage location.

\subsection{Shake Table Experiment 1}

In this and next subsections, we will validate the proposed algorithm using the data collected from two shake table experiments, both were conducted at National Center for Research on Earthquake Engineering (NCREE) in Taiwan. In experiment 1, two identical three-story steel frames were installed on the same shake table. Specimen 2 had a weakened column on the first floor, resulting in a loss of stiffness, as indicated in Fig.~\ref{fig:taiwan1}. The white noise excitation with $0.05g$ amplitude was applied in the North-South direction. More details on this experiment are presented in \cite{liao2015application}. For this analysis, we use the responses collected by the accelerometers installed on the Northwest columns. We assume the responses from the Specimen 1 and Specimen 2 represent the pre-damage condition and post-damage conditions respectively. The data are processed with the same procedure as the damage detection section. \rev{The normalized data are modeled as an AR model with an order of $p=7$, which is selected by using AIC values.} All seven AR coefficients are employed as the DSFs. 

\begin{figure}
\centering
\includegraphics[width=0.8\linewidth]{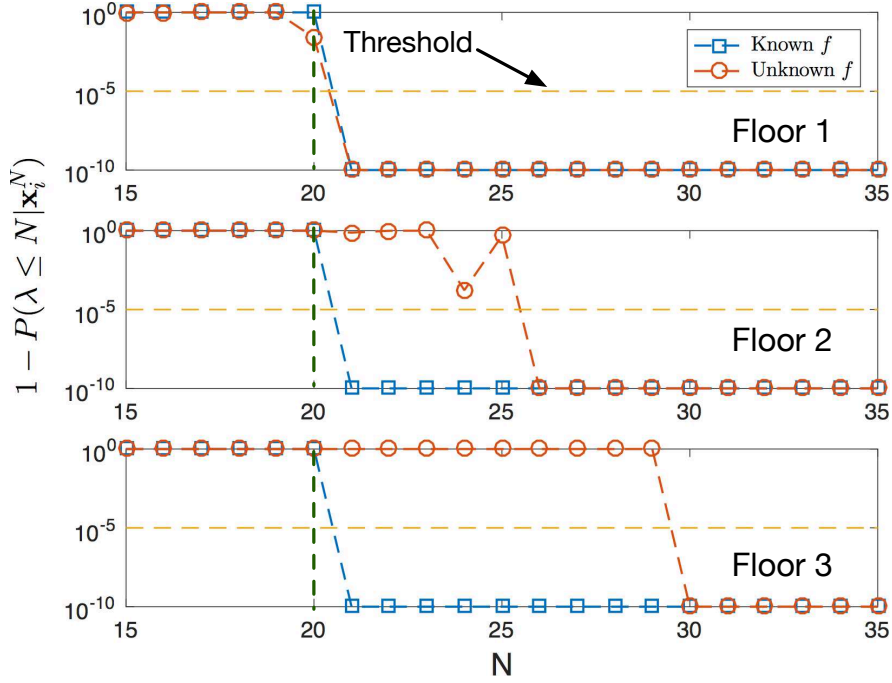}
\caption{Experiment 1: CCDF of each floor.}
\label{fig:tw1_postlike_plot}
\end{figure}

Fig.~\ref{fig:tw1_postlike_plot} shows the complementary cumulative distribution function (CCDF), $1-P(\lambda \leq N|\mathbf{x}_i^N)$, of all sensors, with and without knowing the parameters of $f(\mathbf{X})$. \rev{When the damage does not occur, the cumulative distribution function (CDF) $P(\lambda \leq N|\mathbf{x}_i^N)$, which describes the probability of damage occurrence, is small. Once the damage occurs, the CDF starts to increase and therefore, CCDF starts to decrease. When the fault exists for a long time, the CDF is close to one and CCDF is close to zero.} The damage occurrence time is $\lambda = 20$, as indicated by the vertical dashed line in Fig.~\ref{fig:tw1_postlike_plot}. When the CCDF drops below the threshold (horizontal dashed line), we declare a damage event. In Fig.~\ref{fig:tw1_postlike_plot}, all sensors detect damage with no latency when the distribution parameters are known because the KL distances are significant large for all sensors. Although $\text{DI}_2$ has limited performance in this test, fortunately, we can still use $\text{DI}_1$ to localize damage. Specifically, Sensor 1 has the largest $D_{KL}(f(\mathbf{X})\|g(\mathbf{X}))=305.1590$. The KL distances of Sensor 2 and 3 are $207.2592$ and $138.1704$. When the parameters of $f(\mathbf{X})$ are unknown, Sensor 1, which is close to the damage location, reports damage with no delay. \rev{For Sensor 2 and 3, as shown in Fig.~\ref{fig:tw1_postlike_plot}, compared with Sensor 1, they need more time steps to declare a fault. However, Sensor 2 is still faster than Sensor 3 to identify damage.} Hence, both $\text{DI}_1$ and $\text{DI}_2$ can identify damage location when $f(\mathbf{X})$ is unknown.



\subsection{Shake Table Experiment 2}
The damage patterns in the simulation and experiment 1 are considered as major damage. To study the performance of our algorithm on the minor and moderate damage patterns, we use the data collected from a different shake table experiment. \rev{In shake table experiment 2, we firstly study the column yielding, which is considered as \textit{minor} damage pattern. Then, we replace the bolts of joint with weaken ones and study the joint failure, which is considered as \textit{moderate} damage pattern.} Fig.~\ref{fig:taiwan2} illustrates the diagram of the structure, which is a four-story steel frame. To introduce structural damage, the testing specimen was subjected to a series of the scaled 1999 Taiwan Chi-Chi earthquake. The peaks of strong motions were increased progressively from $0.1g$ to $0.9g$. A white noise excitation with a amplitude of $0.05g$ was conducted between two strong motion excitations. All excitations were applied along North-South direction. All 18 accelerometers were sampled by a common Analog-to-Digital Converter (ADC) with a sampling frequency of $200Hz$. The acceleration response from each white noise excitation is fitted with a 7th order AR model and all coefficients are used as DSF. The parameters of $g(\mathbf{X})$ are estimated using the data from the white noise excitation before any shaking. 




\begin{figure}
\centering
\includegraphics[width=1.1\linewidth]{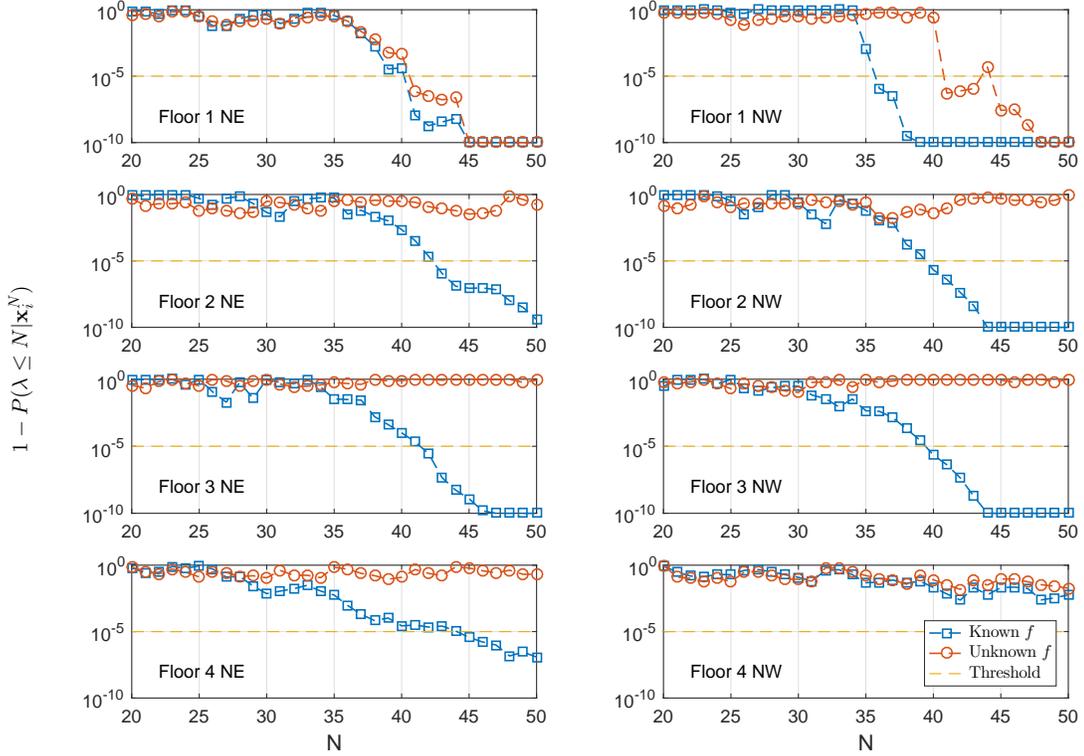}
\caption{Experiment 2 (yielding): CCDF of each floor. $\lambda=35$.}
\label{fig:tw2_postlike_plot_yielding}
\end{figure}

After the strong motion excitation with an amplitude of $0.3g$, several yielding points were observed on all columns of first floor, which are considered as a minor damage pattern. Fig.~\ref{fig:tw2_postlike_plot_yielding} illustrates the CCDFs of all sensors that measure vibration along the North-South direction. The damage (yielding) occurs at $\lambda=35$. The complementary probabilities of the 1st floor sensors immediately drops at $N=35$ when $f(\mathbf{X})$ is known. However, compared with the major damage pattern (Fig.~\ref{fig:tw1_postlike_plot}), the detection delay is longer. For example, the sensor on NE side of floor 1 has a delay of six time steps. The reason is that the KL distance in the minor damage pattern is smaller than that in the major damage pattern. Therefore, according to Lemma~\ref{thm:opt_delay}, the proposed detector needs more data to declare a damage event for the same $\alpha$. Compared with sensors on other floors, both sensors on the floor 1 still have the quickest detection. Therefore, both $\text{DI}_1$ and $\text{DI}_2$ can be used as the localization indices. When $f(\mathbf{X})$ is unknown, the sensors on the floor 1 still detect damage. Also, the additional detection delay is within a reasonable range. For the sensor on NE side of floor 1, the additional delay is zero. For the sensor on NW side of floor 1, the additional time step is $4$. Since the chunk size is $1600$ data points, each time step is equivalent to $8$ seconds and the additional detection delay is $32$ seconds, which is still reasonable for the minor damage pattern.

\begin{figure}
\centering
\includegraphics[width=\linewidth]{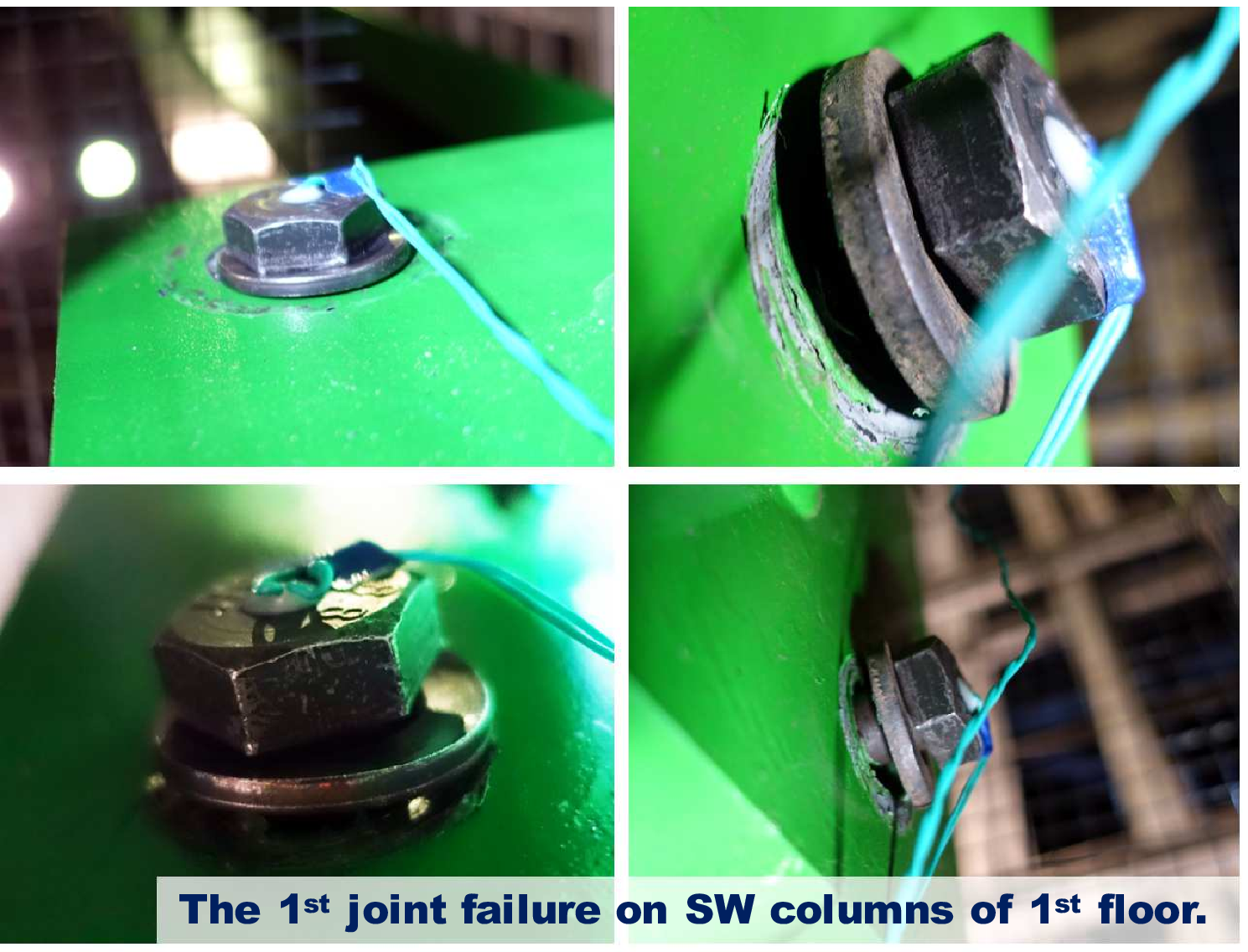}
\caption{Experiment 2: washers and bolts after $0.7$g strong motion excitation.}
\label{fig:joint_failure}
\end{figure}

\begin{figure}
\centering
\includegraphics[width=\linewidth]{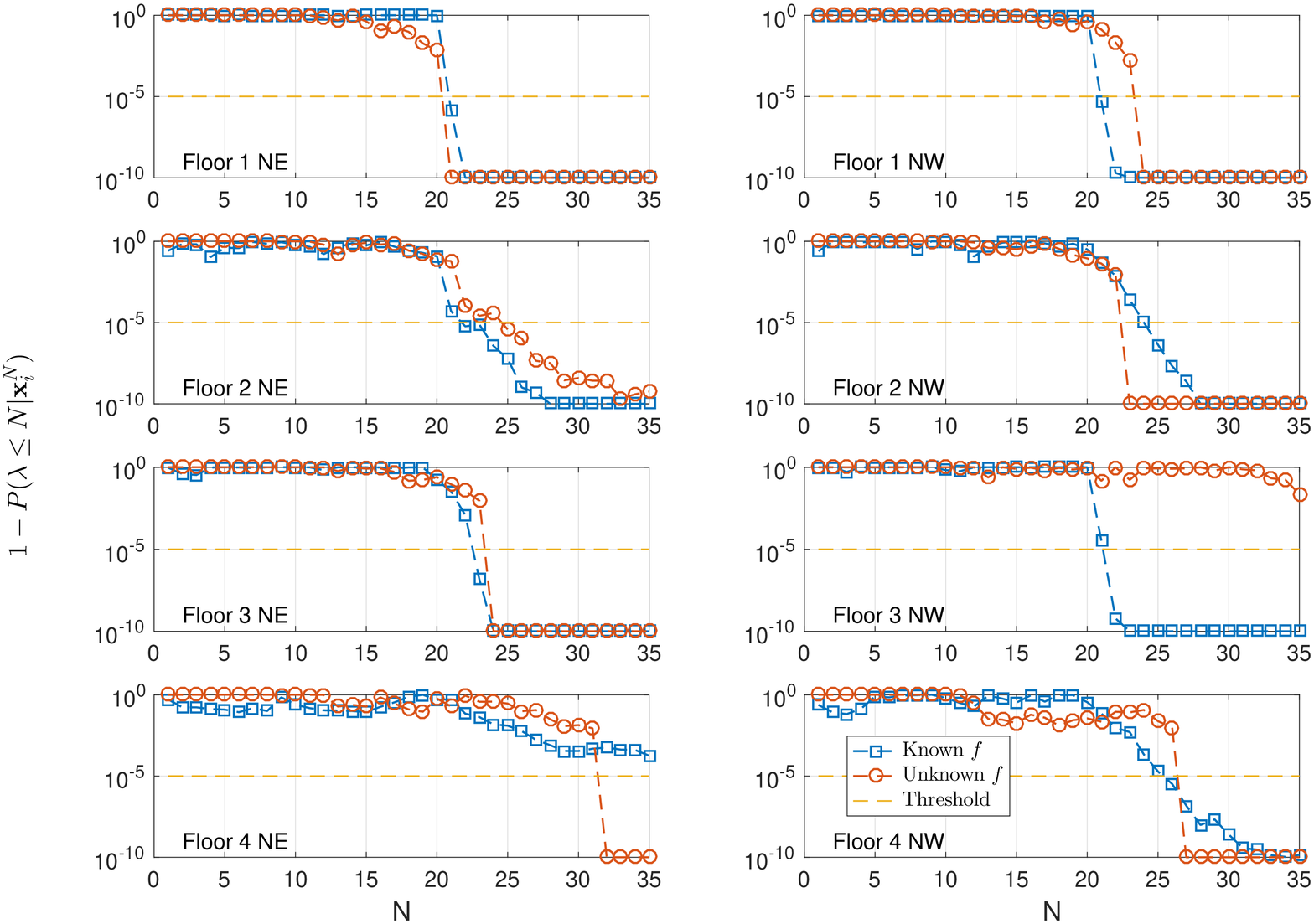}
\caption{Experiment 2 (joint failure, N-S direction sensors): CCDF of each floor. $\lambda = 21$.}
\label{fig:tw2_joint_x}
\end{figure}

For moderate damage patterns, we replaced the joint of SW column of the 1st floor with lossy bolts. After the strong motion excitation of $0.7g$, the washers and bolts became loose, as shown in Fig.~\ref{fig:joint_failure}. Fig.~\ref{fig:tw2_joint_x} shows the CCDFs of all sensors that measure vibration along the North-South direction. The damage occurrence time is $\lambda = 21$. When $f(\mathbf{X})$ is known, both sensors on the first floor has no detection delay. For sensors on other floors, they need more time to descent below the threshold due to small KL distances. Therefore, we can use $\text{DI}_1$ and $\text{DI}_2$ to localize damage. When $f(\mathbf{X})$ is unknown, the trends of the CCDF are similar to the cases when the $f(\mathbf{X})$ is known but more data are required to declare a damage event.

\begin{figure}
\centering
\includegraphics[width=\linewidth]{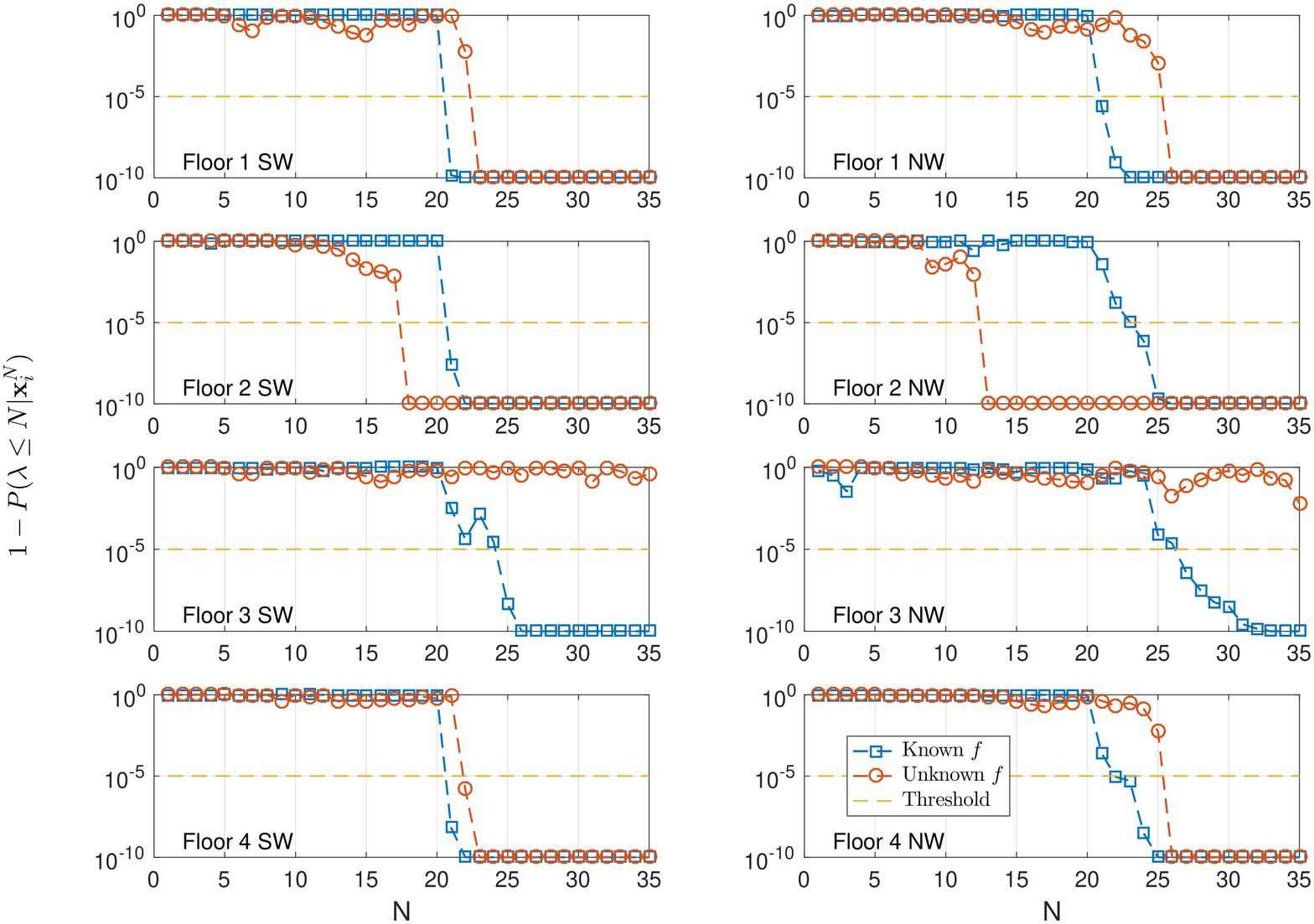}
\caption{Experiment 2 (joint failure, E-W direction sensors): CCDF of each floor. $\lambda = 21$.}
\label{fig:tw2_joint_y}
\end{figure}

When a joint fails, the structure becomes asymmetric and rotation happens. Therefore, to find the damage location precisely, we also use the data of the sensors along East-West direction. In Fig.~\ref{fig:tw2_joint_y}, when $g(\mathbf{X})$ is known, the CCDFs of sensors on the first floor reduces below the threshold without any detection delay. Also, all sensors on the South-West column also report damage quickly. With the localization indices of North-South direction sensors, we can accurately find the damage location as the South-West column of 1st floor. When the parameters of distribution $f(\mathbf{X})$ are unknown, the sensors on the 2nd floor have false alarms. The sensor on the SW column of 4th floor reports damage earlier than the sensor on the same column of 1st floor. But when we use the parameter estimates to compute the KL distances of each sensor, the KL distances of sensors on the 1st floor are the largest. Therefore, we can still accurately find the damage location by using $\text{DI}_1$.

\section{Conclusions}
\label{sec:con}
In this paper, we present a statistical algorithm for automatically detecting and localizing structural damage in a sequential manner. Specifically, we propose a change point detection approach based on the DSF probability distribution changes because of damage events. As a highlight, unlike existing methods, our method does not require the damage pattern as a prior. In addition to damage detection, we directly utilize the detection results to find damage locations. For validation, we applied this algorithm to the simulated data set of the ASCE benchmark structure and the data sets collected from two shake table experiments. The results demonstrate that when the post-damage information is missing, the proposed damage detectors still have optimal performance. Also, the proposed damage localization indices are accurate and sensitive to damage location, with known and unknown post-damage feature distribution.

\section{Appendix: Unknown Post-Damage Distribution Parameter Estimation}
\label{sec:par_est}
In this section, we will provide details on how to find (\ref{eq:parm_est}). Since $g(\mathbf{X})$ and $f(\mathbf{X})$ are Gaussian distributions, (\ref{eq:approx_log_post_prob}) can be written as
\begin{eqnarray*}
	\widetilde{P}(\lambda \leq N | \mathbf{x}_i^N) &=& \log C + \sum_{k=1}^N-\frac{\pi(k)}{2}\left(\sum_{n=1}^{k-1}\log|2\pi\Sigma_0|+(\mv_i[n]-\boldsymbol{\mu}_0)^T\Sigma_0^{-1}(\mv_i[n]-\boldsymbol{\mu}_0)\right. \nonumber \\
	&+& \left. \sum_{n=k}^{N}\log|2\pi\Sigma_1|+(\mv_i[n]-\boldsymbol{\mu}_1)^T\Sigma_1^{-1}(\mv_i[n]-\boldsymbol{\mu}_1)\right).
\end{eqnarray*}
To estimate $\boldsymbol{\mu}_1$, we have
\[
	\frac{\partial \widetilde{P}(\lambda \leq N | \mathbf{x}_i^N)}{\partial \boldsymbol{\mu}_1} = \sum_{k=1}^N-\frac{\pi(k)}{2}\sum_{n=k}^{N}(\mv_i[n]-\boldsymbol{\mu}_1)\Sigma_1^{-1} = 0.
\]
Therefore,
\begin{eqnarray*}
	\sum_{k=1}^N\pi(k)\sum_{n=k}^{N}(\mv_i[n]-\boldsymbol{\mu}_1) &=& \sum_{k=1}^N\pi(k)\left(\sum_{n=k}^{N}\mv_i[n] - (N-k+1)\boldsymbol{\mu}_1\right) = 0 \\
	\widehat{\boldsymbol{\mu}}_1 &=& \frac{\sum_{k=1}^N\pi(k)\sum_{n=k}^{N}\mv_i[n]}{\sum_{k=1}^N\pi(k)(N-k+1)}.
\end{eqnarray*}
Similarly, for the covariance matrix $\Sigma_1$, we have
\[
\frac{\partial \widetilde{P}(\lambda \leq N | \mathbf{x}_i^N)}{\partial \boldsymbol{\Sigma}_1} = \sum_{k=1}^N-\frac{\pi(k)}{2}[(N-k+1)\text{tr}(\Sigma_1^{-1}(\partial\Sigma_1))-\text{tr}(\Sigma_1^{-1}(\partial\Sigma_1)^{-1})\Sigma_1^{-1}S_k],
\]
where $S_i[k] = \sum_{n=k}^N(\mv_i[n]-\boldsymbol{\mu}_1)(\mv_i[n]-\boldsymbol{\mu}_1)^T$. If we let $\boldsymbol{\mu}_1 = \widehat{\boldsymbol{\mu}}_1$, the covariance matrix estimate is 
\[
\widehat{\Sigma}_1 = \frac{\sum_{k=1}^N\pi(k)S_i[k]}{\sum_{k=1}^N\pi(k)(N-k+1)}.
\]

\section{Acknowledgement}
We would like to thank Shieh-Kung Huang from National Taiwan University (NTU) and personnel of NCREE for their help and collaboration. This research is partially supported by the NSF-NEESR Grant 1207911 and their support is gratefully acknowledged. The first author would like to thank the Charles H. Leavell Graduate Student Fellowship for the financial support.

%
%

%

\end{document}